\documentclass[%
 aip,
 sd,%
 amsmath,amssymb,
 reprint,%
]{revtex4-1}

\usepackage{graphicx}% Include figure files
\usepackage{dcolumn}% Align table columns on decimal point
\usepackage{bm}% bold math
\usepackage{mathtools}
\usepackage{xcolor}
\usepackage{graphicx}
\begin{document}

\preprint{AIP/123-QED}

\title{Effect of Inhomogeneous magnetic field on Plasma generation in a low magnetic field helicon discharge}% Force line breaks with \\
%\thanks{Footnote to title of article.}
\author{Sonu Yadav}
\email{syadav@ipr.res.in}
\affiliation{Institute for Plasma Research, HBNI, Bhat, Gandhinagar 382428, India}%Lines break automatically or can be forced with \\

\author{Prabal K Chattopadhyay}
\email{pkchatto@ipr.res.in}
\affiliation{Institute for Plasma Research, HBNI, Bhat, Gandhinagar 382428, India}%Lines break automatically or can be forced with \\

\author{Kshitish K. Barada}
 \affiliation{Department of Physics and Astronomy, University of California, Los Angeles, CA 90095, USA}%Lines break automatically or can be forced with \\

\author{Soumen Ghosh}
 \affiliation{Department of Physics, University of California, San Diego, CA 92093, USA}%Lines break automatically or can be forced with \\
 \author{Joydeep Ghosh}

\affiliation{Institute for Plasma Research, HBNI, Bhat, Gandhinagar 382428, India}%Lines break automatically or can be forced with \\

\date{\today}

\begin{abstract}
The ionization efficiency of helicon plasma discharge is explored by changing the low axial magnetic field gradients
near the helicon antenna. The highest plasma density is found for a most possible diverging field near the antenna 
by keeping the other operating condition constant. Measurement of axial wave number together with estimated radial
wavenumber suggests the oblique mode propagation of helicon wave along the resonance cone boundary. Propagation of 
helicon wave near the resonance cone angle boundary can excite electrostatic fluctuations which subsequently can 
deposit energy in the plasma. This process has been shown to be responsible for peaking in density in low field 
helicon discharges, where the helicon wave propagates at an angle with respect to the applied uniform magnetic field.
The increased efficiency can be explained on the basis of multiple resonances for multimode excitation by the
helicon antenna due to the availability of a broad range of magnetic field values in the near field of the antenna 
when a diverging magnetic field is applied in the source.   
%
%Valid PACS numbers may be entered using the \verb+\pacs{#1}+ command.
\end{abstract}

% \keywords{Hollow density, magnetically expanding plasma, off-axis additional ionization}%Use showkeys class option if keyword

\maketitle
\section{\label{sec:level}Introduction:}
Efficient plasma sources are studied for their application as an economic plasma source for micro-electronics
industry \cite{ref_1}, negative ion production \cite{ref_2}, neutron generation \cite{ref_3} and space propulsion \cite{ref_4}. 
The helicon plasma source is one of the most efficient plasma source. This plasma source is well known for
their generation of high density plasma at relatively low powers and magnetic field. Beyond the certain
values of rf power and magnetic field, it is found that the helicon wave can be absorbed in the plasma 
and generate the high density. A large number of experiments have been conduct in various tube length 
and antenna geometry.  Normally helicon sources operate at a magnetic field which corresponds to an electron 
cyclotron frequency 20 to 40 times of source frequency. This means for normal operation of the helicon source 
using 13.56 MHz source 200 G magnetic field or more is required. However, helicon sources have shown resonance 
abortion even around a particular low magnetic field. Here, low magnetic field (20-30 G) means $f_{ce} \sim$  5 times 
source frequency. 

Recently, low magnetic field helicon sources are proposed for space propulsion \cite{ref_5,ref_6,ref_7} and also can be used 
as sources for semiconductor applications \cite{ref_8} as they can drastically reduce the cost of the magnet power 
supply and subsequently the power consumption. Resonance absorption at low magnetic
fields \cite{ref_9,ref_10,ref_11,ref_12,ref_13,ref_14,ref_15,ref_16,ref_17} less
than 100 G in a helicon discharge when a magnetic field is increased contrary to its behavior of monotonic
increase at higher magnetic fields \cite{ref_18}. All these experiments are carried out using uniform magnetic fields 
near the antenna. However,  Lafleur et al \cite{ref_17} using a diverging magnetic field using one coil only near the
antenna have reported a ten-fold increase in density compared to the density at zero magnetic field. While
using two coils they observed that the density does not increase as much compared to while using only the
single coil. In that reported experiment, the source coil current is varied and the current in the exhaust 
coil is kept constant at one of the values between 0-5 A (in 1 A steps). With the exhaust coil kept on, 
the non-uniformity of the magnetic field near the antenna decreases. No explanations are given for higher 
density production for the non-uniform magnetic field Case. Our earlier results \cite{ref_19} on the density peaking 
phenomena in a uniform low magnetic field near antenna have shown that density increase is due to the 
resonance behavior while wave propagating near resonance cone. Signature of excitation of low frequency 
electrostatic fluctuations supports this hypothesis. As the helicon modes (radial eigenmodes) in a bounded 
system are discrete, only a single mode can undergo this phenomenon at a critical magnetic field when a 
uniform magnetic field is applied near the antenna. For a non-uniform magnetic field, there could be the 
resonance of different radial eigenmodes. This might lead to higher production efficiency. Hence, study the 
density peaking behavior in the nonuniform magnetic fields is useful. 

In the present work, plasma production efficiency is studied by applying different diverging magnetic 
field configuration near the source (antenna). The plasma density increases with increasing the magnetic field non-uniformity
near the antenna. It is observed that the plasma density is highest at particular non-uniform magnetic
field configuration at low B in our experiments. These results may be thought to be due multimode cyclotron
resonance due to the nonuniform magnetic fields near the antenna.

This paper is organized as follows. 
The experimental setup and diagnostics are described in section II followed by experimental results
and discussion in section III and summary is presented in section IV.

\section{\label{sec:level} Experimental Setup and Diagnostics:}
\begin{figure}
\centering
\includegraphics[width = \linewidth]{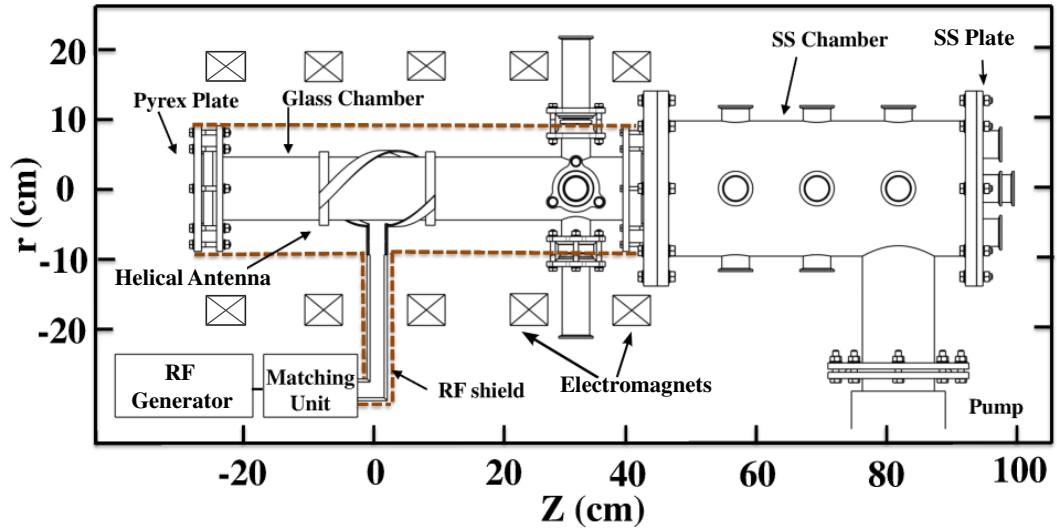}
\caption{Schematic diagram of the Helicon plasma experimental setup} \label{fig:1}
\end{figure}

The experiments are performed in the linear helicon plasma experimental device ( Fig. \ref{fig:1}), which has 
been described before \cite{ref_20,ref_21}. The vacuum system consists of a source and expansion chamber. 
The source chamber is made of borosilicate glass tube and has 9.5 cm inner diameter and 70 cm length and 
the left end is terminated by a insulating plate. The other end of the source tube is attached to a 51 cm 
long stainless steel (SS) expansion chamber of 21 cm inner diameter. The whole chamber is evacuated to a 
base pressure of $1\times10^{-6}$ mbar by a diffusion pump connected to the expansion chamber. 
The Argon is used as the working gas and experiments are carried out at pressure $1\times10^{-3}$ mbar. 
An 18 cm long, m = +1 right helicon antenna, placed around the source chamber is used to produce plasma 
by means of a 13.56 MHz rf power generator through an L-type impedance matching network. 
The reflected power is kept less than 2$\%$ for all the experiments. 
The axial location of the antenna center is defined as z = 0 and all other axial locations are with
reference to this antenna center as shown in Fig. \ref{fig:1}.

\begin{figure}
\centering
\includegraphics[width = 0.8\linewidth]{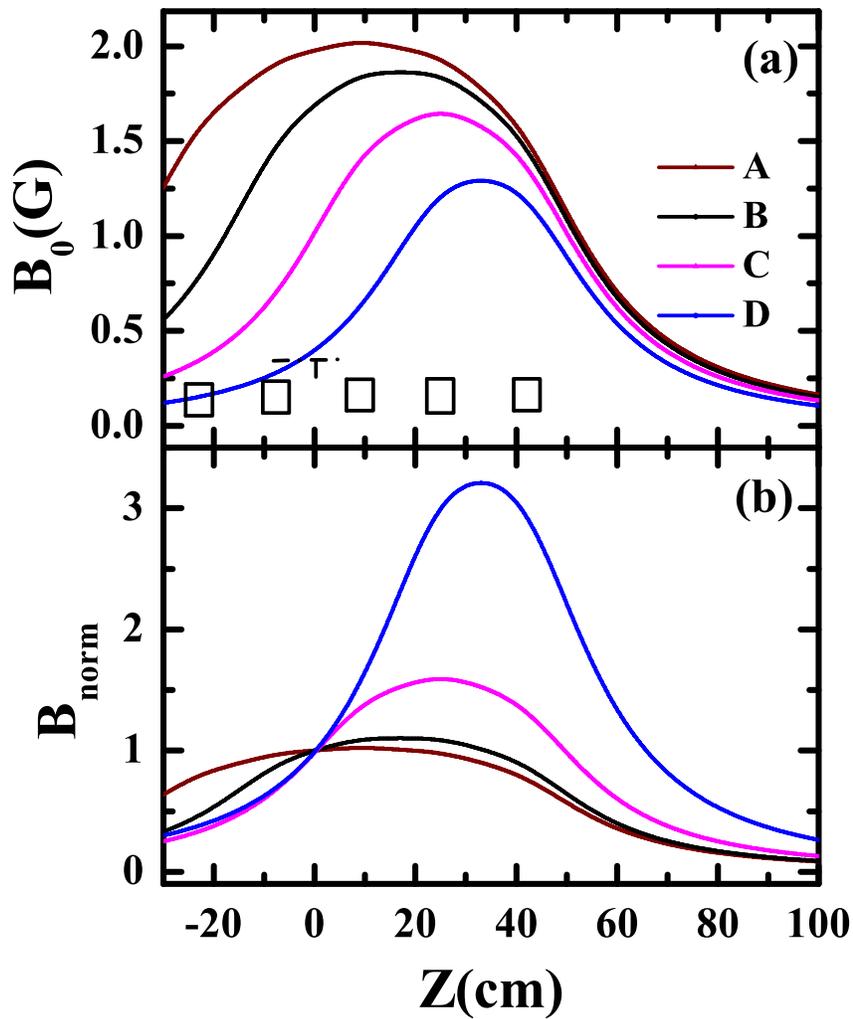}
\caption{Axial variation of (a) magnetic field strength and (b) Normalized magnetic field with respect to 
field value at antenna center for Cases of A-D when 1 A current is passed in the coils} \label{fig:2}
\end{figure}

Five water-cooled electromagnet coils, as shown in  Fig. \ref{fig:1}, are used to generate an axial magnetic field ($B_0$).
Four different magnetic field configurations are used in our experiments, produced by passing direct current 
(DC) through different electromagnet combinations. Fig. \ref{fig:2}a shows the axial variation of the 
simulated magnetic field when 1A DC current is passed through different electromagnet combinations 
to produce four different magnetic field configurations. Fig. \ref{fig:2}b shows the axial variation of 
normalized magnetic field with respect to field value at antenna center. Fig. \ref{fig:2}b noticeably 
indicates the magnetic field homogeneity or inhomogeneity in each configurations. The leftmost coil 
as shown in the Fig. \ref{fig:1}, is the first coil and the rightmost coil is the fifth coil. Case A corresponds 
to a magnetic field profile when 1 A current is passed through all 5 coils (coil 1-5). 
The Case B magnetic field is produced by taking out the first coil and keeping the other 4 (coil 2-5).
The Case C magnetic field is obtained by removing the first and second coils from left and
keeping the other 3 coils (coil 3-5). Case D magnetic field corresponds to 1 A current in the 
fourth and last coil by removing the first three coils from the left. 

Two separate single rf compensated Langmuir probes \cite{ref_21}, one straight, inserted from the top radial port in the
source chamber at z = 31 cm and other L shaped, inserted off-axially from the end flange of the expansion
chamber at z = 50 cm, are used to measure the plasma density, $n_0$. These locations also correspond to before
($z_{before}$ = 31 cm) and after ($z_{after}$ = 50 cm) both in the magnetic and geometric expansion Fig. \ref{fig:1}. 
The probe tips of nearly equal collection area, are made up of cylindrical tungsten wire of
1 mm diameter and 4 mm length. The plasma density determined from the ion saturation current by 
considering the sheath expansion \cite{ref_22}.

The wave detection is performed by measuring the amplitude and phase of axial wave magnetic field ($B_z$), 
by positioning the high frequency B-dot probe about 10 cm downstream from the antenna center and pulling 
it back in 2 cm increments. The B-dot probe is a single loop of 6 mm diameter made from the shielded coaxial 
cable of 1.8 mm outer diameter. At each position, the amplitude is measured as the peak-to-peak value 
averaged over ten shots. To measure the axial phase variation, the antenna current measured simultaneously
using a high frequency Rogowski coil, acts as a phase reference. Both signals 
(antenna current and B-dot probe) are fed to separate channels of a digital oscilloscope. 

\begin{figure}
\centering
\includegraphics[width = 0.8\linewidth]{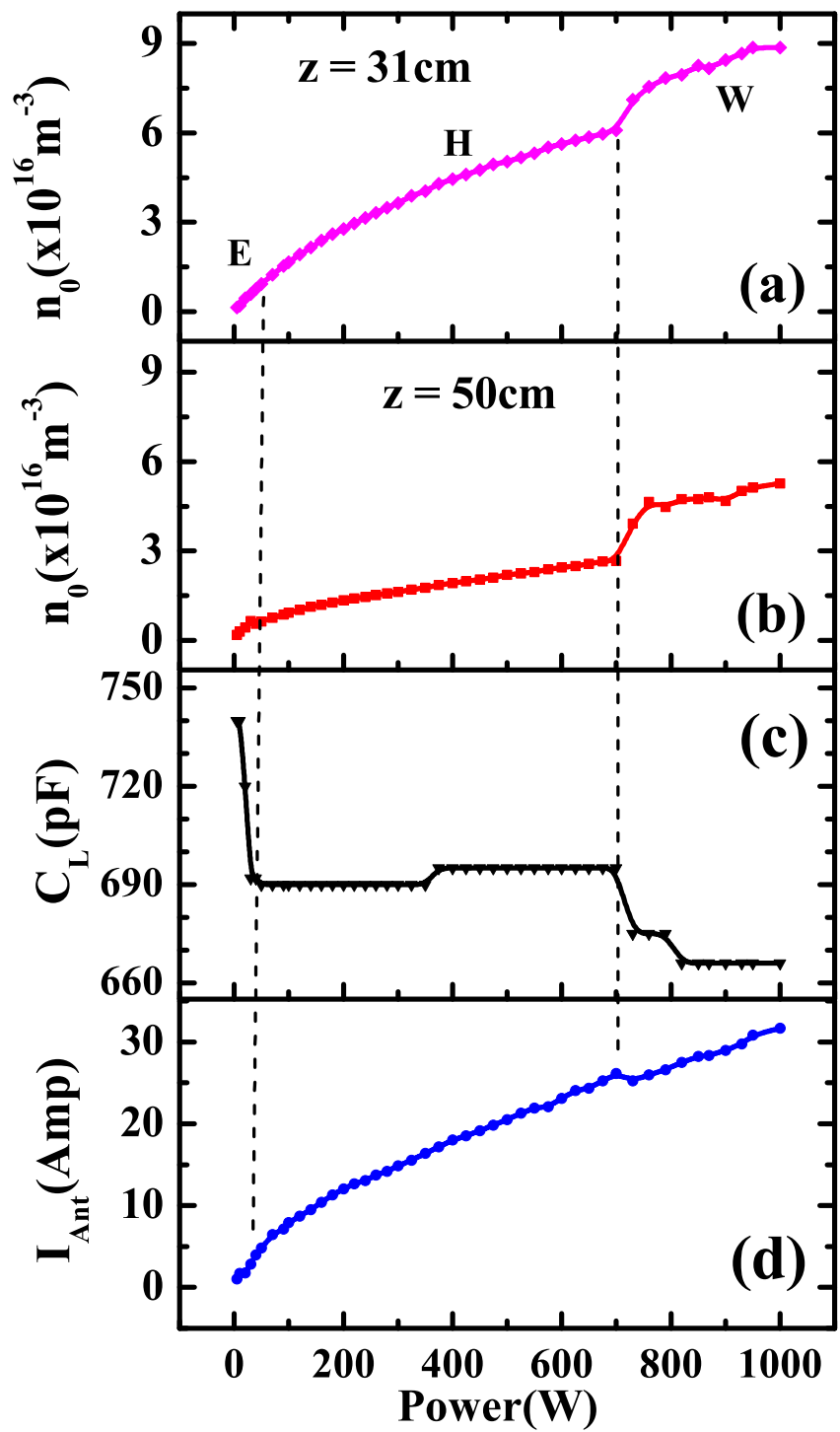}
\caption{Helicon discharge mode transitions. Langmuir probe measurement of center plasma density at
(a) z = 31cm, (b) z = 50cm and (c) Value of load capacitor ($C_L$) and (d) Antenna rms current as function of rf power,  
measured at $1\times10^{-3}$ mbar and $B_0$ = 160G ($I_B$ = 87A). Dash lines show the 
capacitive (E) to inductive (H) and inductive to helicon wave (W) mode transitions.} \label{fig:3}
\end{figure}

\begin{figure}
\centering
\includegraphics[width= 8cm]{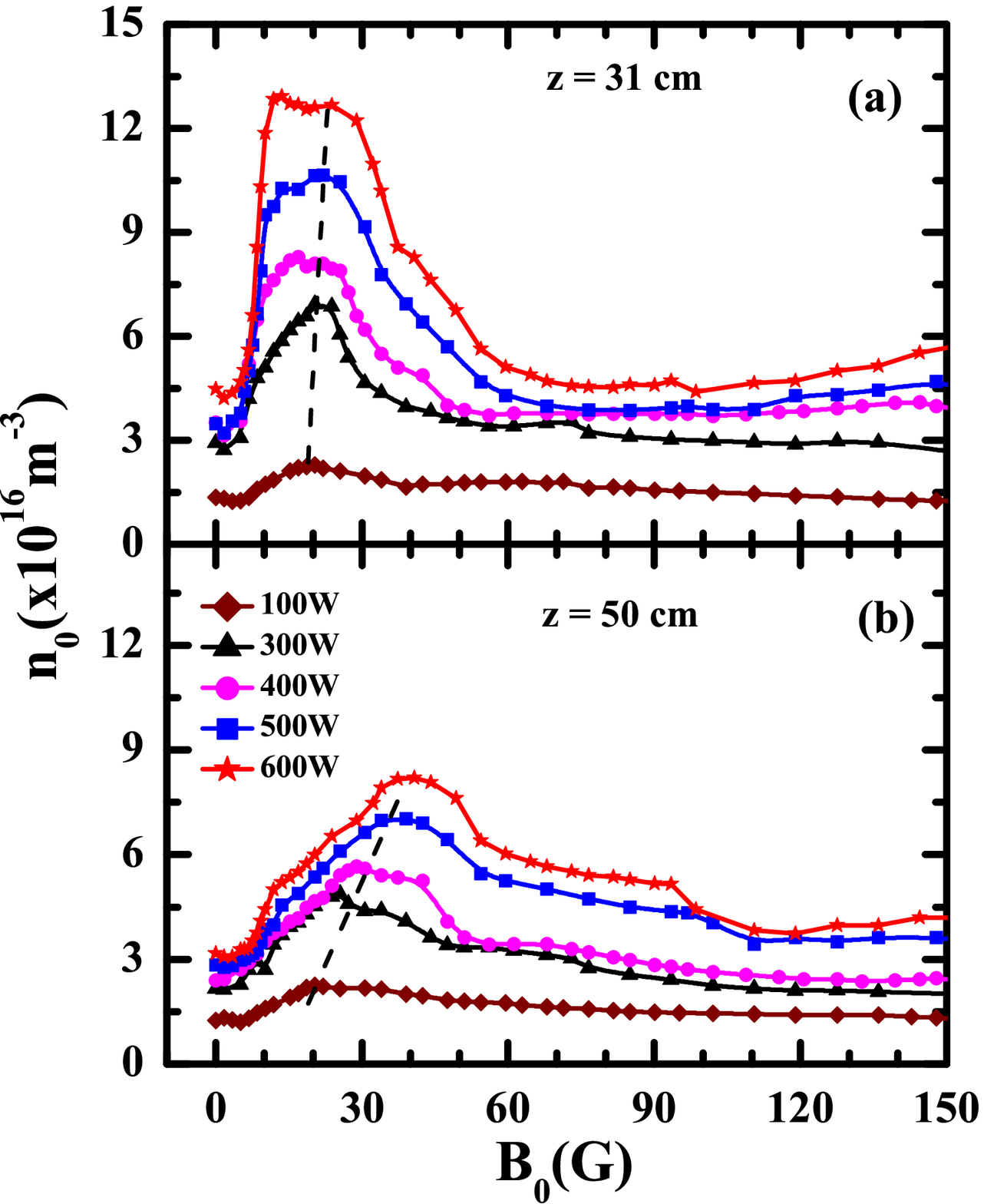}
\caption{On-axis variation of plasma density with magnetic field at antenna center for the Case B in the
(a) source chamber at z = 31cm and (b) expansion chamber at z = 50cm, for different values of rf powers,
100W solid diamonds, 300W solid triangles, 400W solid circle, 500W solid square and 600W solid star. }\label{fig:4}
\end{figure}

\section{\label {sec:level} Experimental Results and Discussion}
It is well know that the helicon discharges are quite versatile and depending on operating parameters,
the discharge can be operated in capacitive (E), inductive (H) or Helicon wave (W) mode \cite{ref_1,ref_9,ref_23}.
To determine the different discharge mode in our experiment, we measured the center plasma density
in both source chamber at z = 31cm (Fig. \ref{fig:3}a) and expansion chambers at z = 50cm (Fig. \ref{fig:3}b). 
The $n_0$ is measured in quasi-uniform magnetic field configuration, i.e. Case B as function of rf
input power at $1\times10^{-3}$ mbar pressure and B0 $\sim$ 160G ($I_B$ = 87A) applied magnetic field. Fig. \ref{fig:3} shows the
gradual and abrupt changes in center plasma density that can be identified as E-H and H-W transitions 
near 50 and 700W respectively. The discharge mode transitions are also associated with discrete change 
in values of load capacitor ($C_L$), Fig. \ref{fig:3}c with optimized reflected power less than 2$\%$ by keeping
the fixed value of tune capacitor ($C_T$) in matching network \cite{ref_20}. The measured values of rms antenna current 
($I_{ant}$) as function of rf power is shown Fig. \ref{fig:3}d, it also reflects the mode transitions near 50 and 700W. 
It is found that discharge mode transition from H to W is clearly observed in source and expansion plasma 
density for power 700W.  The density jump at B0 $\sim$ 160G  is associated with the helicon wave abortion \cite{ref_23}.

In our previous experimental work \cite{ref_19} it has been shown that the helicon wave is also absorbed even at low magnetic 
field ($<$ 100G). The multiple density peaks at low magnetic fields has already been observed in our Helicon Experimental 
device \cite{ref_20} when the uniform magnetic field is applied near the antenna in the source chamber. In the present experiments, 
the phenomenon of density peaking at the low magnetic field is explored by changing the axial magnetic field gradient
at the antenna location. The $n_0$ measurement is performed in the four different axial magnetic field 
configurations. These different magnetic field configurations are entitled by the Case A, B, C and D as shown in Fig. \ref{fig:2}. 
The $n_0$ is obtained using the rf compensated Langmuir probe at two different axial locations, z = 31 cm and 50 cm. 
For all present Cases (A, B, C, and D) the values of the magnetic field at abscissa are taken at antenna center i.e. z = 0 cm. 

Fig. \ref{fig:4}a and Fig. \ref{fig:4}b show the $n_0$ as function of magnetic field for the Case B for different
rf powers at z = 31cm and z = 50 cm
respectively. As the magnetic field is slowly increased from 0 G, the density rises monotonically till 15 G (at 300 W)
to 25 G (at 600W), before falling sharply to about half of its peak value near 50-60 G in the source chamber, at z = 31 cm 
(Fig. \ref{fig:4}a). The effect of low magnetic field phenomena is also present in the bigger expansion chamber, Fig. 3b. 
The density rise in the expansion chamber (far away from the antenna) is monotonic till 25-40 G but fall is rather gradual 
than source chamber. The fall of density with magnetic field is not same in both the chambers may be due different 
diffusion process involved. At low rf power (100W) the very weak density peak is present in both the source and
expansion chambers. The magnitude of the peak density is higher and the profile becomes broader for increased rf powers.
It is also noted that with increasing rf power the density peak shifted to larger magnetic field values.
The similar observations can also be found in the previous experiments \cite{ref_15,ref_24,ref_25}. 

\begin{figure}
\centering
\includegraphics[width=0.8\linewidth]{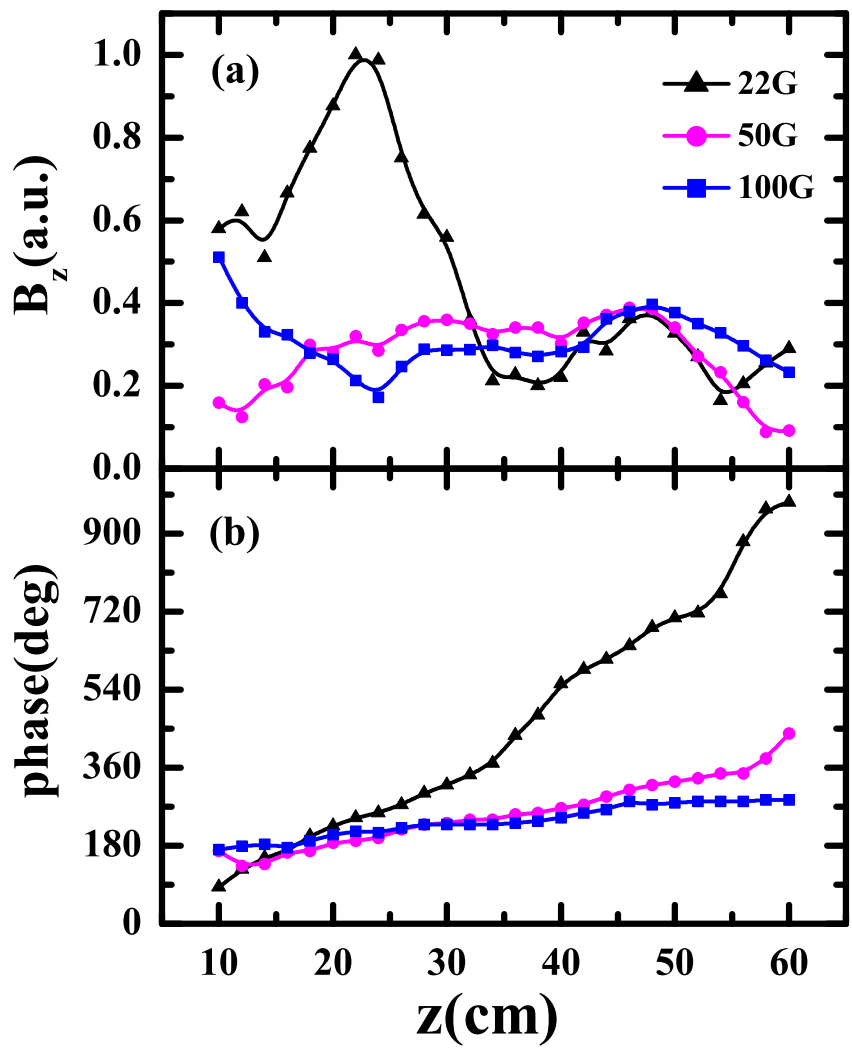}
\caption{Axial variation of (a) amplitude and (b) phase of an axial component of wave magnetic field ($B_z$) at 400W rf power
and $1\times10^{-3}$ mbar for the Case B at magnetic field value 22 G, solid triangles, 50 G solid circle and 100 G solid square.}
\label{fig:5}
\end{figure}

It has already been established that the density peaking at the low magnetic field is the results of wave coupling
rather than capacitive or inductive coupling \cite{ref_17,ref_19}. In reference 19 it was shown that the wave propagation near the 
resonance cone surface causing the density peaking. To establish the existence of helicon wave, the measurement of phase
and amplitude of axial wave magnetic field $B_z$, are carried out using the B-dot probe. Fig. \ref{fig:5} shows the axial
variation of
the phase and amplitude of $B_z$ at 400W rf power in the Case B. At 22 G, the axial variation of wave amplitude,
Bz (Fig. \ref{fig:5}a) has a spatial modulation whereas, phase difference (Fig. \ref{fig:5}b) shows the traveling wave character. 
This type of  amplitude modulation can be explained by the beating of different helicon wave modes corresponding
to fundamental and higher order radial modes, simultaneously, excited by the antenna \cite{ref_11,ref_26,ref_27}.
These different helicon modes are propagating with different angles of the axial magnetic field. 
Depending on the magnetic field strength near the antenna, one of these modes can predominantly 
take part in the ionization process by resonant absorption.  An effective traveling wavelength can be found by the slope
$\frac{d \phi}{dz}$ of fitted straight line to axial phase variation (Fig. \ref{fig:5}b), where $\phi$ is the phase difference.
At 22 G, in the near field of the antenna the effective traveling wavelength, ${\lambda_\parallel}$ = 360$(\frac{d \phi}{dz})^{-1}$
, is about 34 cm. However, for the high magnetic field at 50 and 100 G, nearly constant phase variation (Fig. \ref{fig:5}b)
suggested the barely existence of helicon wave. The amplitudes also does not indicate any wave pattern, 
(Fig. \ref{fig:5}a). This implies that the non-existence of a helicon wave at 50 and 100G in our experimental conditions 
in the case B.  

The angle of obliquely propagating helicon wave with the axial magnetic field can be found from the relation, 
$cos\theta = \frac{k_\parallel}{k}$, where $k = \sqrt{k_\parallel^2 + k_\perp^2}$, $k_\parallel$ and $k_\perp$ are the 
total, parallel and perpendicular wavenumbers, respectively. 
The calculated value of wave propagation angle is comes about 76.40$^\circ$ by using the measured values of parallel wavenumber
, i.e. $k_\parallel$ = 18.4 $m^{-1}$ and estimated perpendicular wavenumber $k_\perp {_1}$. The value of
$k_\perp {_1}$ = $\frac{3.83}{a}$
= 76.6 $m^{-1}$ can be found from the $1^{st}$ root of the first order Bessel function $J_1(k_\perp a)$  
, by considering the plasma radius $a$ = 5 cm \cite{ref_28}.
The calculated value of resonance cone angle \cite{ref_19}, $cos\theta = \frac{\omega}{\omega_{ce}}$, for 22 G 
at which the density peak occurs (Fig. \ref{fig:4}a), is comes about 77.30$^\circ$. It is found that the helicon wave 
propagation angle well matches with the resonance cone angle. 
The details of the relation between resonance cone angle and wave propagation angle is described in reference 19. 

\begin{figure}
\centering
\includegraphics[width=0.8\linewidth]{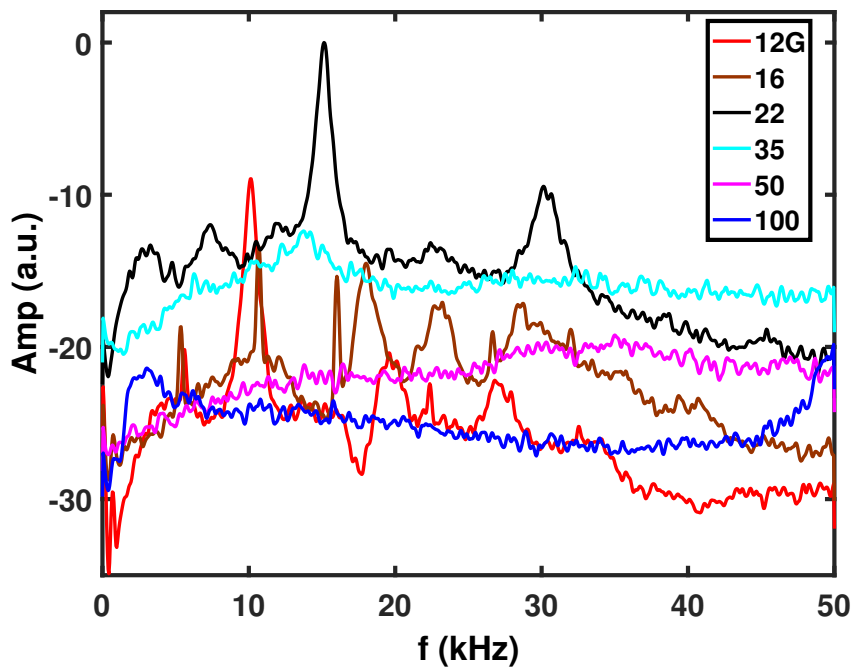}
\caption{The frequency spectrum of plasma density fluctuation obtain at the different magnetic field with rf
compensated Langmuir probe in Case B at on-axis location of z = 31cm, 400W rf power and $1\times10^{-3}$ mbar fill pressure. } \label{fig:6}
\end{figure}

The absorption of helicon wave near the resonance cone surface is correlated with the excitation of electrostatic 
fluctuation \cite{ref_29}. In the present experiment, the measurements of density fluctuation are performed at the magnetic
field values where helicon wave present and absent. For this, an electrostatic probe (rf compensation Langmuir probe)
biased in the ion saturation region, is placed in the source chamber at $r,z = 0,31$ cm. A 14-bit PXI based data
acquisition system is used to acquire the time series signal with low pass filter of 1.9 MHz with 50k record length 
at the sampling rate of 100 kS/s. The frequency spectrum (FFT) of density fluctuation at 400 W rf power and $1\times10^{-3}$ mbar
fill pressure for various magnetic field strength in the Case B is shown in Fig. \ref{fig:6}. 
The strong peaks are observed between 10 to 20 kHz frequency at 22 G of magnetic field value, above this
there is no significant peak in the frequency spectrum. The presence of electrostatic fluctuations at
low magnetic field ($\sim$ 25G) are associated with the density peaking (Fig. \ref{fig:4}) where helicon wave also exist. 
The magnetic field $>$ 25 G, the non-existence electrostatic fluctuations is consistent with non-propagation 
of helicon wave (Fig. \ref{fig:5}) and hence this can be correlated to density fall Fig. \ref{fig:4}.

\begin{figure}
\centering
\includegraphics[width=0.8\linewidth]{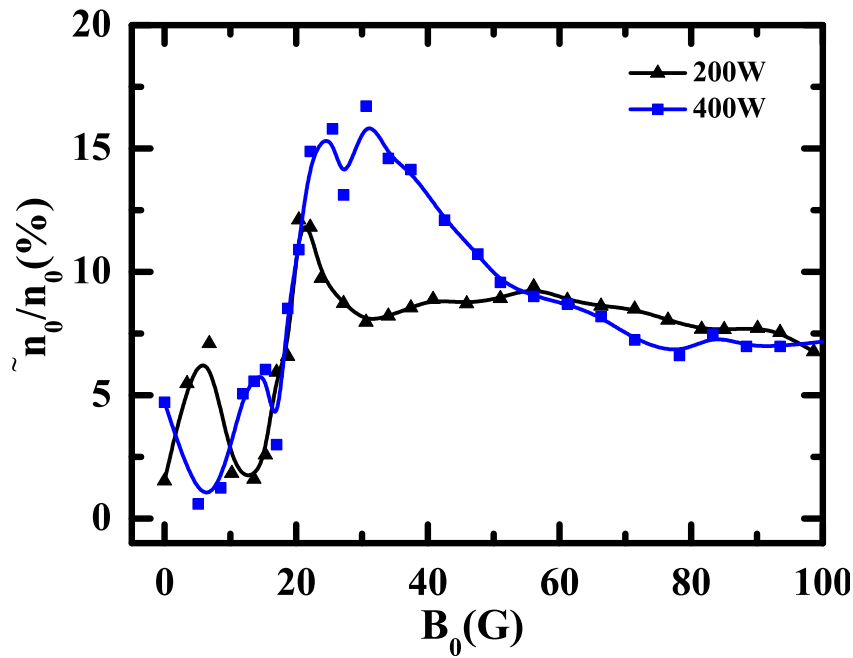}
\caption{Amplitude variation of density fluctuation with the external magnetic field as a function of rf powers in Case 
B at the location of $r,z = 0,31$cm and $1\times10^{-3}$ mbar fill pressure. 200 W solid triangles and 400 W solid square.}
\label{fig:7}
\end{figure}

Fig. \ref{fig:7} shows the variation of density fluctuation level with an applied magnetic field for two different powers 
at pressure $1\times10^{-3}$ mbar. It is observed that level of fluctuation increases significantly at those magnetic fields where 
the density peaks occur. Therefore in our experiments, the oblique helicon wave propagation near the resonance cone 
boundary excites the non-convective electrostatic wave \cite{ref_19,ref_29,ref_30,ref_31} which transfers the energy from helicon wave to
the plasma creating density peaks at specific magnetic field. This process is responsible for density peaking at
the low magnetic field in our experiments.

\begin{figure}
\centering
\includegraphics[width=0.8\linewidth]{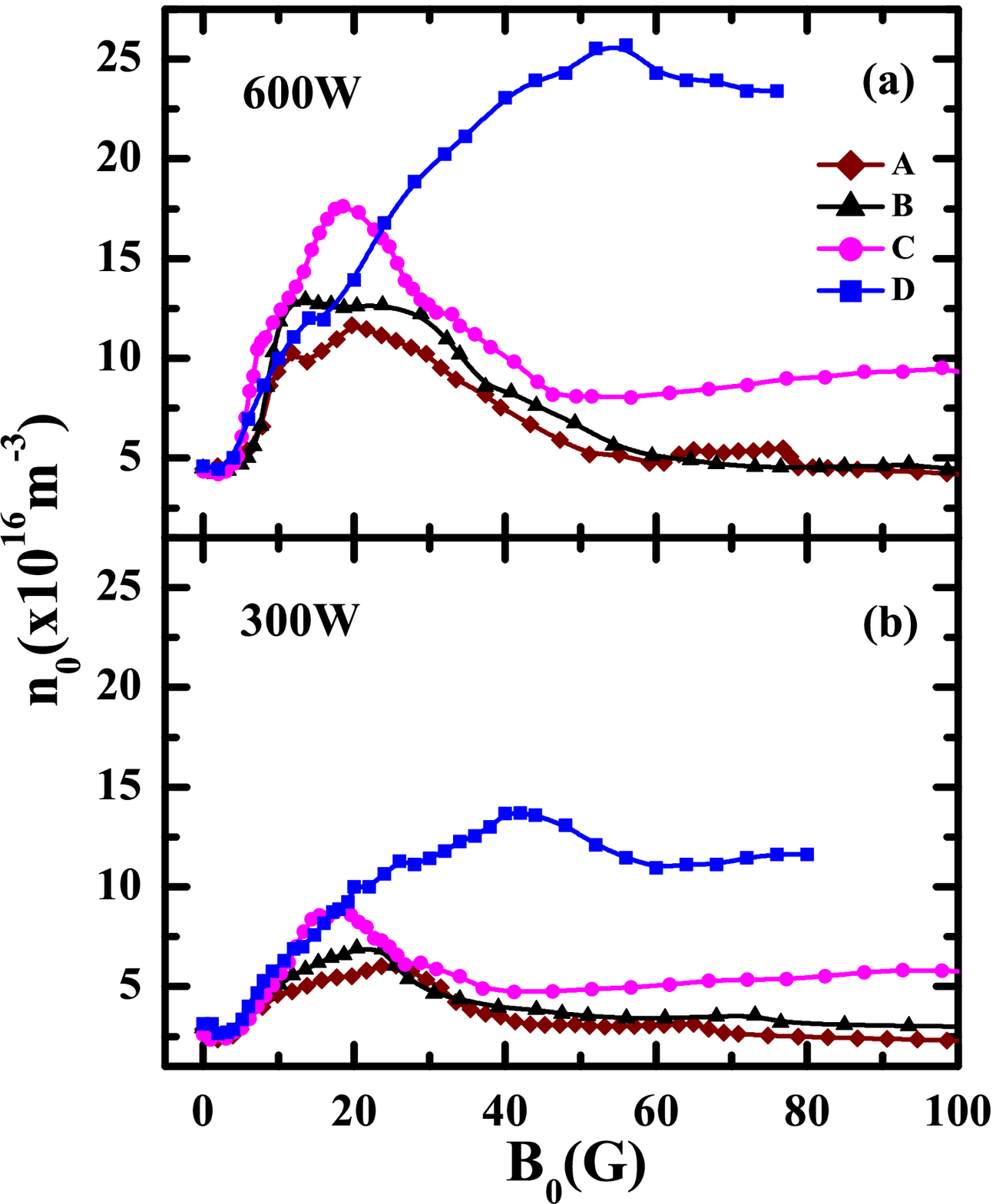}
\caption{Variation of plasma density in the source chamber at z = 31cm with the applied magnetic field at antenna center for 
Cases A-D; at pressure $1\times10^{-3}$ mbar and rf powers (a) 600W and (b) 300W.} \label{fig:8}
\end{figure}

\begin{figure}
\centering
\includegraphics[width=0.8\linewidth]{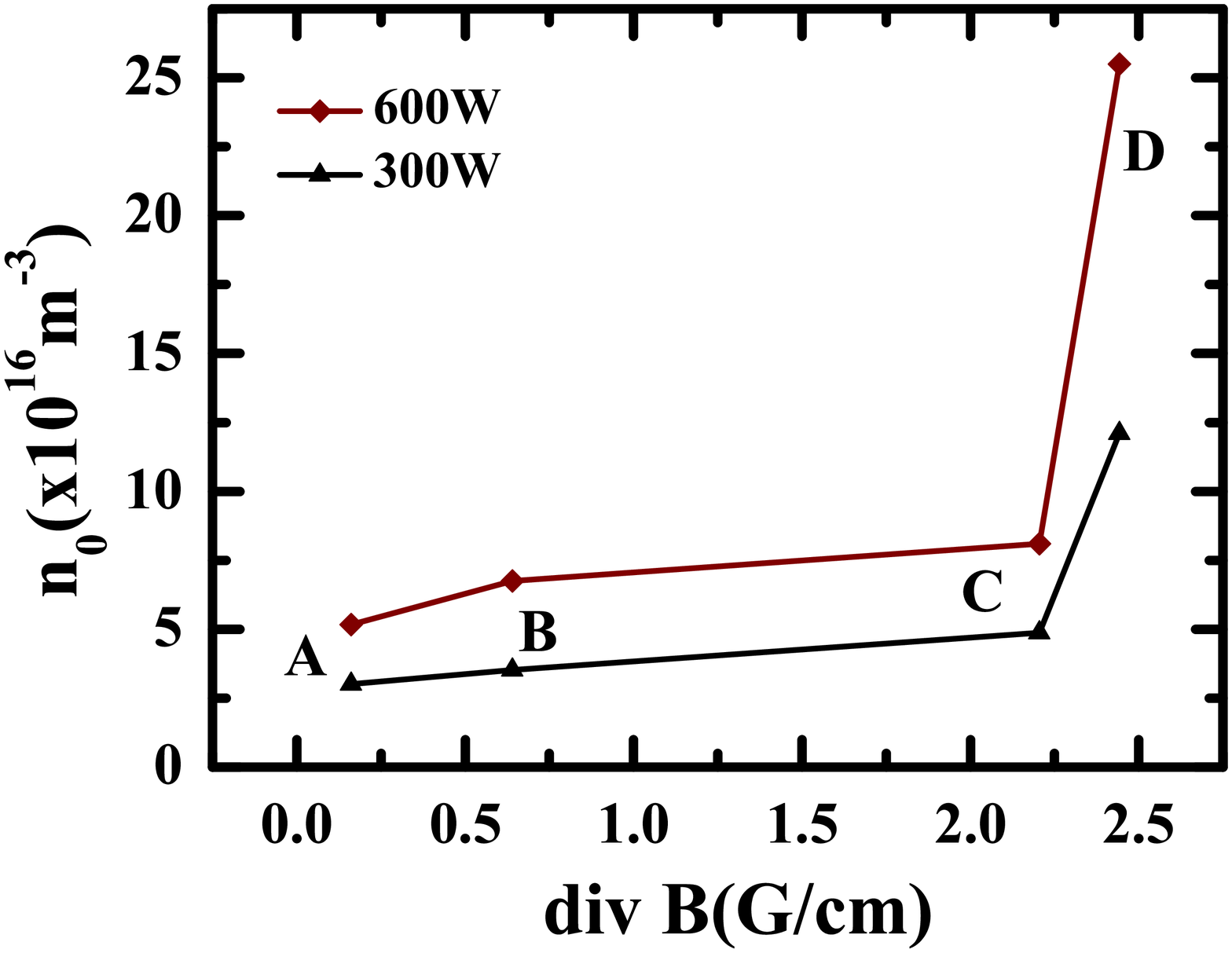}
\caption{Density as function of magnetic field non-uniformity in different Cases at 50G and $1\times10^{-3}$ mbar pressure.}
\label{fig:9}
\end{figure}

\begin{figure}
\centering
\includegraphics[width=0.8\linewidth]{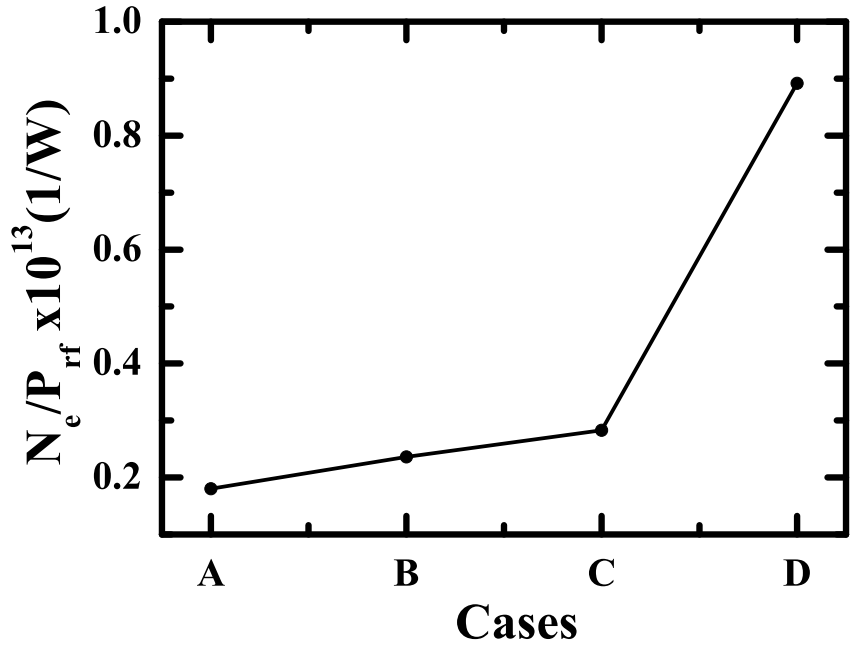}
\caption{On-axis plasma production efficiency at z = 31cm, in different Cases at 50G, 
600W and $1\times10^{-3}$ mbar pressure. }
\label{fig:10}
\end{figure}

After establishing the presence of the helicon wave when density peaked at low magnetic field in the quasi-uniform 
magnetic field configuration, the measurement of plasma density in the source chamber is made for other different 
magnetic field configuration as shown in Fig. \ref{fig:2} by Case A, B, C, and D. In all the Cases the magnetic field is varied
and density is measured at the location, z = 31 cm at $1\times10^{-3}$ mbar pressure at 600 W (Fig. \ref{fig:8}a) and
300 W (Fig. \ref{fig:8}b). 
To get the same magnetic field at the antenna center (z = 0), different current is passed for different coil configurations.
As the coils are removed, it can be seen from Fig. \ref{fig:2}, that field non-uniformity is increases from
Case A to Case D. For Cases A, B and C the density is peaked around 17-22 G and the peak value increases as 
non-uniformity increases for the Case A to C. With removing the one more coil from the case C, the magnetic
field near the antenna become more non-uniform or diverging in the Case D. In the case D, the density peaked 
around ~ 50-60 G and increases about five times than the uniform magnetic field case at the same magnetic field value.
Fig. \ref{fig:9} shows the density at 50G, increases as magnetic field non-uniformity increases from case A to D. 
The plasma production efficiency, \cite{ref_32} $\frac{N_e}{p_{rf}}$ which is the ratio of number of total electrons 
to input power is also increases with the magnetic field inhomogeneity. Fig. \ref{fig:10} shows the production
efficiency estimated at on-axis at z = 31 cm in all four cases. To estimate the production efficiency, 
the production region is important, i.e. plasma core instead of plasma edge. It is seen from above discussion that 
the inhomogeneous magnetic field near the antenna location improves the plasma production significantly.    

Our results show a systematic variation of plasma density with magnetic field inhomogeneity.
Plasma production is more when the magnetic field is more diverging near the antenna.
This kind of behavior of increase of production efficiency with magnetic field divergence may be attributed
to multimode oblique cyclotron resonance. In the direction of the magnetic field, the m = +1 right helicon antenna 
is supposed to excite different axial and radial eigenmodes \cite{ref_27, ref_33}. Depending upon the magnetic field near the
antenna all these different modes can simultaneously couple to different helicon modes and couple power in the plasma.
In our earlier experiments \cite{ref_19} 
it was shown that multiple modes can be excited at different uniform magnetic fields near the antenna. 
These modes can have oblique cyclotron resonance for a critical magnetic field which satisfies
$cos\theta = \frac{\omega}{\omega_{ce}} = \frac{k_{\parallel}}{k} = \frac{k_{\parallel}}{\sqrt{k_{\parallel}^2 + k_{\perp}^2}}$.
If multiple modes are present, they will have multiple resonances at different critical magnetic fields. 
With a uniform magnetic field, only a single resonance is possible for a mode. When a non-uniform magnetic field 
(because of the divergence) is applied, different magnetic field values are available near the antenna near
field for different eigenmodes resonance conditions. Among all the possible excited modes some modes can have resonances.
This will lead to more power absorption and hence rise in the efficiency of the source as observed in the present experiment.
Another interesting feature to be noted here is that the width of the peaks of density increases as the 
non-uniformity in the magnetic field is increased. This is indicative of the presence of different resonating modes. 
As low field helicon sources are proposed recently for space propulsion \cite{ref_5,ref_6}, using a highly diverging magnetic field
near the antenna may be better suited in terms of efficiency of the source. It will reduce the weight of the
thrusters as well as the power consumption. 

\begin{figure}
\centering
\includegraphics[width=0.8\linewidth]{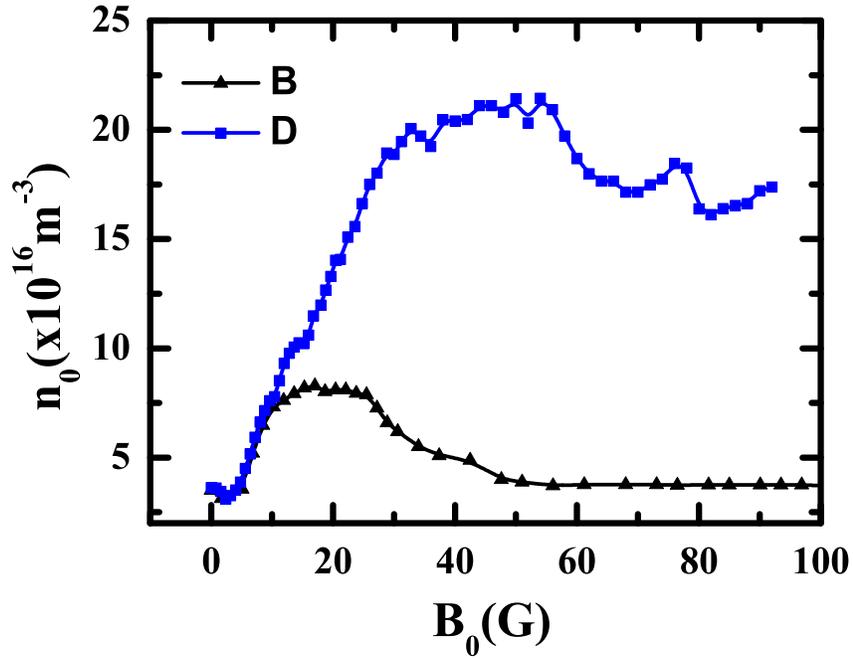}
\caption{On-axis variation of plasma density with the magnetic field at 400 W RF power and $1\times10^{-3}$ mbar 
fill pressure. Case B (solid triangle) and Case D (solid square).}
\label{fig:11}
\end{figure}

\begin{figure}
\centering
\includegraphics[width=\linewidth]{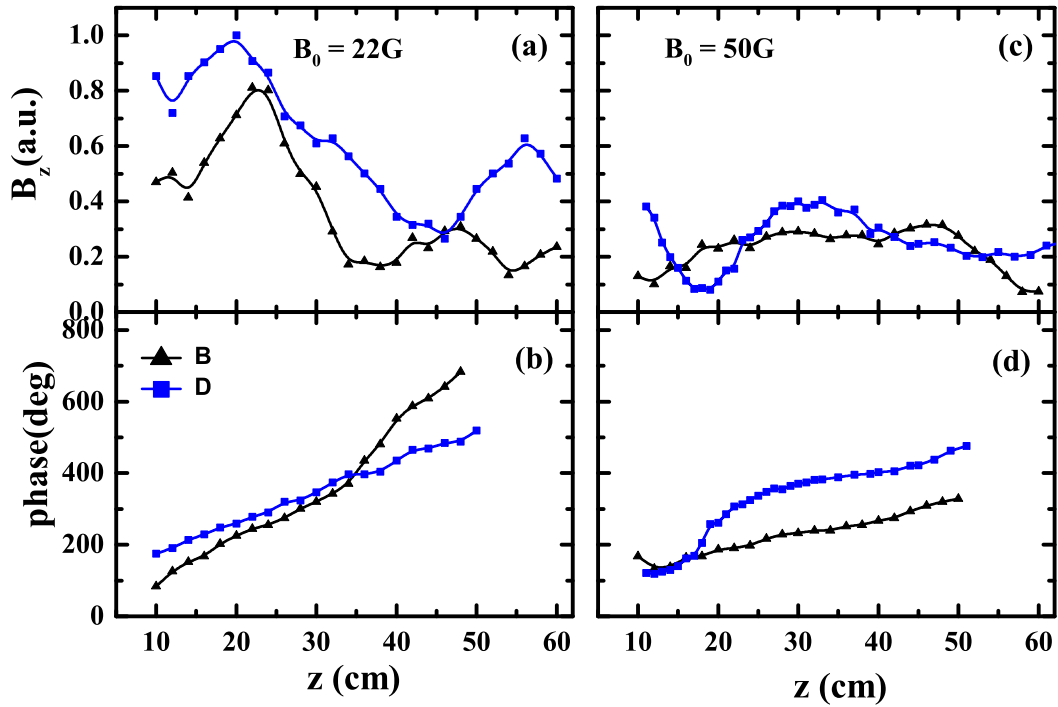}
\caption{Axial variation of amplitude and phase of $B_z$ for Case B and D magnetic field configuration 
for 22 G (a and b) and 50 G (c and d) at 400 W rf power and $1\times10^{-3}$ mbar fill pressure.}
\label{fig:12}
\end{figure}

\begin{figure}
\centering
\includegraphics[width=0.8\linewidth]{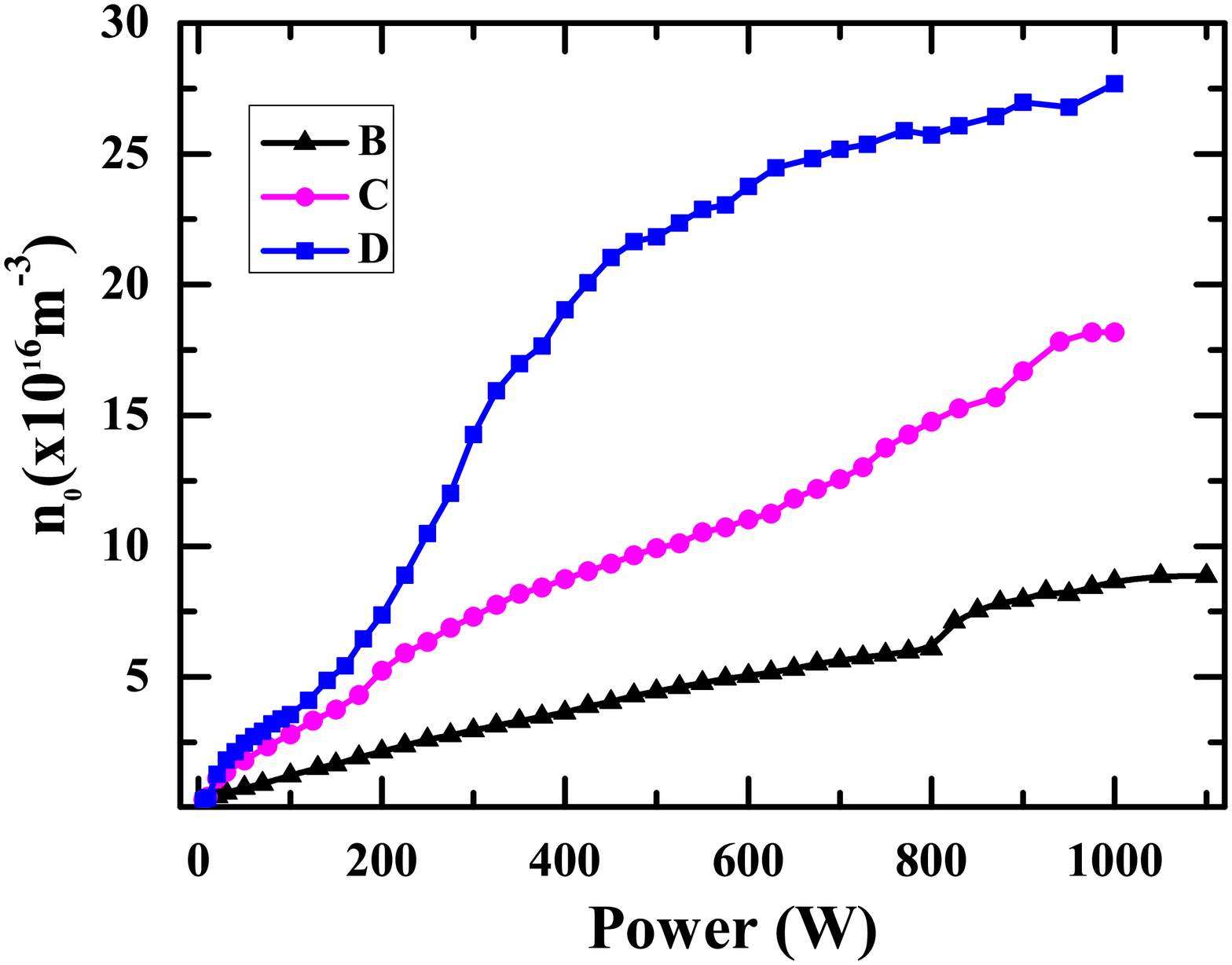}
\caption{Variation of on-axis plasma density at z = 31cm with 87 A DC current for the Cases of B (solid triangle), 
C (solid circle) and D (solid square) at pressure $1\times10^{-3}$ mbar.}
\label{fig:13}
\end{figure}

Fig. \ref{fig:11} shows the on-axis plasma density for the Case B and D at 400 W rf power. 
The plasma density is peaked around 40-50 G in the Case D and it is nearly 6 times higher than the Case B 
at that magnetic field value.  
To investigate the presence of multimode in the inhomogeneous magnetic field configuration,
the axial profiles of the wave amplitude and phase of $B_z$ are measured in the Case D and comparisons are made 
with the quasi-uniform magnetic field configuration, i.e. Case B. Fig. \ref{fig:11} shows the density variation with magnetic 
field for Case B and D at which wave amplitude and phase of $B_z$ are measured. Fig. \ref{fig:12} shows the axial variation of the 
amplitude (Fig. \ref{fig:12}a and c) and the phase difference (Fig. \ref{fig:12}b and d) of $B_z$ at two different
magnetic field values,
22 G ($\frac{\omega_{ce}}{\omega_{rf}}$ = 4.5) and 50 G ($\frac{\omega_{ce}}{\omega_{rf}}$ = 10),
respectively for the Case B (solid triangle) and D (solid square) at 400 W rf power and
$1\times10^{-3}$ mbar fill in pressure. At 22 G, for both cases B and D, the phase difference variations (Fig. \ref{fig:12}a) show
the traveling wave character and amplitude (Fig. \ref{fig:12}b) modulations show the beating wave pattern. 
The beating pattern \cite{ref_11,ref_26,ref_27} of amplitude in presence of propagating wave indicates the existence 
of more than one (radial or axial) mode in our experiment. At 50 G, for Case B (where density has 5 times 
lower magnitude from the Case D, Fig. \ref{fig:11}), the phase difference has no significant axial variation 
(Fig. \ref{fig:12}d, solid triangles) i.e. wave propagation may ceases at this magnetic field value. 
The contrasting results are observed in the Case D at 50 G (where the density is peaked), 
where $B_z$ has amplitude modulation (Fig. \ref{fig:12}c, solid square) and the phase oscillates (Fig. \ref{fig:12}d, solid square)
substantially with the position. It seems that, in the Case D, helicon wave exist along the entire axial 
distance including the expansion chamber and has partially traveling and partially standing wave structure.
This implies that waves can have the interference of multiple radial or axial modes. 
This results in a standing wave field structure \cite{ref_27,ref_34,ref_35} that gives rise to the nearly constant phase 
signal at 25 $<$ z $<$ 45 cm in the Case D, Fig. \ref{fig:12}d.  The multiple modes exist due to
presence of different exciting condition corresponding to availability of broad range of magnetic field
near the antenna, Case D Fig. \ref{fig:2}. 

Fig. \ref{fig:13} shows the variation of on-axis plasma density with RF powers for the Cases of B, C, and D 
at $1\times10^{-3}$  mbar fill pressure. 
The density in the Cases B, C and D is obtain by keeping the fixed DC current, 87 A, which corresponds to magnetic 
field values at the antenna center about 150, 90 and 34 G, respectively. It is clear from the Fig. \ref{fig:13} that, 
above 300W rf powers, in the Case D, the density is five times higher than the Case B. The density at 1000W,
in case B is about $8\times10^{16}$ $m^{-3}$ which can be achievable at 200W in the Case D. These results successfully demonstrate
that the higher ionization can be achieved by careful placing of electromagnet or permanent magnet near the antenna. 
This will effectively reduce the coast of the magnetic power supply and power consumption which in turn enhanced
helicon discharge efficiency
 
\section{\label {sec:level}  summary and Conclusion:}

Four different magnetic field configurations near the helicon antenna are used to look for the effect of 
magnetic field inhomogeneity. It is shown that plasma production efficiency increases with increasing 
the magnetic field inhomogeneity. The results are explained on the basis of multimode excitation due to 
the presence of a wide range of magnetic fields nearby the antenna. The widths of the peaks in density 
increase with non-uniformity of the applied magnetic field. This is indicative of multimode resonance absorption
leading to higher efficiency. As the less coil is more efficient, the cost of the source will be reduced drastically.

\nocite{*}
\bibliography{Low_B_manuscript.bib}

\end {document}